\documentclass [12pt,twoside]{article}           
\usepackage[left,modulo]{lineno}
\usepackage{setspace}
\usepackage{verbatim}

\countdef\arxiv=201

\ifnum\arxiv=0 \input cpreb.tex \fi

\ifnum\arxiv=1

\usepackage{epsfig,times,lscape}
\usepackage[usenames]{color}

    \ifnum\count102=1

    \fi

    \ifnum\count102=2
\fi

\newcommand{\nc}{\newcommand}

     \ifnum\count101=1
\nc{\qI}[1]{\section{{#1}}}
\nc{\qA}[1]{\subsection{{#1}}}
\nc{\qun}[1]{\subsubsection{{#1}}}
\nc{\qa}[1]{\paragraph{{#1}}}

\def\qpar{\vskip 2mm plus 0.2mm minus 0.2mm}
\def\qL{\hfill \break}
     \fi 

      \ifnum\count101=2
 \nc{\qI}[1]{\parindent=0mm \vskip 8mm 
{\centerline{\LARGE \color{red}#1}}\vskip 3mm}
\nc{\qA}[1]{\vskip 2.5mm \noindent 
{{\bf\large\color{blue}  #1}} \vskip 1mm \parindent=0mm}
 \nc{\qun}[1]{\vskip 1mm \noindent {\sl #1 }\quad }

\def\qL{\hfill \break}
\def\qpar{\vskip 2mm plus 0.2mm minus 0.2mm}

      \fi

\def\qth{\vrule height 12pt depth 0pt width 0pt}
\def\qtb{\vrule height 0pt depth 5pt width 0pt}

\def\qtH{\vrule height 20pt depth 0pt width 0pt}

\nc{\qfoot}[1]{\footnote{{#1}}}

      \ifnum\count101=1
\def\qbu{\hfill \par \hskip 6mm $ \bullet $ \hskip 2mm}
\def\qee#1{\hfill \par \hskip 6mm (#1) \hskip 2 mm}
      \fi
      \ifnum\count101=2
\def\qbu{\hfill \par \hskip 4mm $ \bullet $ \hskip 2mm}
\def\qee#1{\hfill \par \hskip 4mm (#1) \hskip 2 mm}
      \fi

     \ifnum\count101=1 
 
     \fi
     \ifnum\count101=2

  \def\qcitb#1{\noindent \hbox to 102mm{\hfill \small #1} \vskip 1mm}
      \fi

 \def\qpages#1{\count102=0{\loop\advance\count102 by 1
 \null \vfill\eject \ifnum\count102<#1 \repeat}}

\def\qth{\vrule height 12pt depth 0pt width 0pt}
\def\qtb{\vrule height 0pt depth 5pt width 0pt}

\def\qv{\vskip 0.1mm plus 0.05mm minus 0.05mm}
\def\qhu{\hskip 0.6mm}
\def\qhv{\hskip 3mm}

\def\qhw{\hskip 1.5mm}
\def\qleg#1#2#3{\noindent {\bf \small #1\qhw}{\small #2\qhw}{\it \small #3}\qv }
\fi

\begin{document}
\thispagestyle{empty}



\markboth{{\sl \hfill  \hfill \protect\phantom{3}}}
        {{\protect\phantom{3}\sl \hfill  \hfill}}

\color{yellow} 
\hrule height 20mm depth 10mm width 170mm 
\color{black}
\vskip -2.2cm 

 \centerline{\bf \Large Translation into any natural language of the
error messages}
\vskip 2mm
 \centerline{\bf \Large generated by any computer program}
\vskip 15mm
\centerline{\large 
Bertrand Roehner$ ^1 $
}

\vskip 8mm
\large

{\bf Abstract}\quad
Since the introduction of the Fortran programming language
some 60 years ago,
there has been little progress in making error messages more
user-friendly. A first step in this direction is to
translate them into the natural language of the students.
In this paper we propose a simple script for Linux systems
which gives word by word translations of error messages.
It works for most programming languages and for
all natural languages.
Understanding the error messages generated by compilers
is a major hurdle for students who are learning programming,
particularly for non-native English speakers.
Teachers and professors will squander
a substantial amount of time translating
the English vocabulary.
By freeing them of this burden,
an automatic translation will make more time available
for the teaching of programming itself.
Moreover, our personal experience showed us
that many students simply do not have the patience to
overcome this obstacle. Not only do they
never become ``fluent'' in programming but many give up
programming altogether.
Whereas programming is a tool which can be useful in many
human activities, e.g. history, genealogy, astronomy, entomology,
in many countries the skill of programming 
remains confined to a narrow fringe of professional programmers.
In all societies, besides professional violinists there are
also amateurs. It should be the same for programming.
Translating the terms used in error messages also gives an
opportunity for explaining them. 
For instance, the words ``operand'' 
or ``unary'' are almost
the same in English and in French, but they need to be explained
because only few students know what they mean. \qL
It is our hope that
once translated and explained the error messages will be seen
by the students as an aid rather than as an obstacle and that
in this way more 
students will enjoy learning and practicing programming.
They should see it as a funny game.

\vskip 5mm
\centerline{\it Version of 23 July 2015. Comments are welcome}

\vskip 3mm
{\normalsize Key-words: compiler, error messages, translation, programming,
world languages.}

\vskip 10mm

{\normalsize 
1: Institute for Theoretical and High Energy Physics (LPTHE),
Sorbonne Universit\'es, University Pierre and Marie Curie (UPMC), 
Paris, France. Email: roehner@lpthe.jussieu.fr.
}

\vfill\eject

\qI{Motivation}

Just in order to show that error messages are not always easy
to understand, let me mention that while writing this paper
I came across the following error
message: ``Unary operator expected''. Although hardly a beginner
in programming, I had never before seen the word ``unary''.
Obviously, it was not a translation issue because the word
is almost the same in English and in French, but rather a
question of meaning. That is why it is important to say
that our
purpose is not only to translate but also to explain. In the
present case the explanation is fairly simple: an unary
operator expects only {\it one} argument.
\qpar

The purpose of the present paper is to provide a procedure
for translating (and explaining if necessary) 
error messages generated in English into the natural 
language $ Z $ preferred by the
programmer. The  English and $ Z $ versions will be displayed
one under another in the command shell
so that the $ Z $ version can provide an aid even
if it is only a word by word translation.

\qA{Objections}

When we started this project almost all the colleagues to whom
we explained it raised objections and advised us to drop
it. In this flow of objections the most common arguments
were the following. 
\qbu All professional programmers, whatever their country 
of origin, know
English. In other words, translating will be unnecessary and
a waste of time.
\qbu Understanding the meaning of some error messages may prove
difficult even for native English speakers. This shows that 
it is not the natural language which is the main problem
but rather the programming language.
\qbu Error messages have been  in English ever since
the first programming languages were introduced in the early
1950s. This did not prevent the rapid development of computer
science and the advent of the Internet Revolution.
\qbu Anyway, said some colleagues,
as the code is written in English, what is the
point in translating only the error messages? Some colleagues asked
us ironically:
``May be, you also
wish to rewrite the code into another language than English?''
\qpar

As always in such discussions, these arguments were not wrong
but they focused on specific aspects of the problem and discarded
many others. Actually, at the root of it, there are two different
conceptions of programming.

\qA{Narrow versus broad conception of programming}

In the narrow conception, programming is reserved for
professional people working for software companies on big projects.
In this conception, programming will concern only
a very small fraction of the population. \qL
A parallel with the practice of music would be a society in which
there are only professional musicians. In such a society there would
be no musicians playing in a municipal band or orchestra, 
no persons singing
in a choir, nobody playing music at home with friends just for personal
satisfaction. Fortunately, except for a few Italian expressions
(e.g. allegro vivace) 
there is no need to know another language
before learning to read a score and play an instrument.
A similar situation should prevail in programming.

\qpar
In the broad conception a programming language is seen 
in a way not very different from a
natural language that is to say as a means
which allows people to interact with each other.
Even a basic
knowledge of Russian can make a visit to Moscow more enjoyable;
similarly even just a working knowledge
of a programming language will allow people to do things in
their own field of interest which would be impossible otherwise.
As an example, a knowledge
of Java will allow historians to write a script which will
search the Internet more effectively than through a manual search.
In other words, instead of having to rely
on standard search engines, they will be able to create
their personal search engines focused on their own needs
and requirements.
\qpar

In the narrow conception the fact that error messages are in
English is of little importance because through their education
and professional training all programmers will soon get 
used to it.\qL
In the broad conception the picture is different. If historians
in Turkey, Japan or China want to use Java 
to create a personal search engine for a
specific study, the fact that first of all they will have 
to learn (or to re-learn)
English will certainly be a deterrent. It is true
that the instructions of the Java code are themselves
in English, but, as will be seen below, there is
a great difference between code instructions and error messages.

\qA{Coping with error messages: a major hurdle for many students}

On average over the past 50 years there have been very little changes
in the formulation and presentation of error messages%
\qfoot{It is true that the introduction of IDE
(Integrated Development Environment) tools has been a progress
but it affected the error messages only indirectly.}%
.
Fifty years ago programming was restricted to a tiny fraction
of the population, mostly scientists for which the
cryptic form of error messages was not a big problem.
In 2015, programming is (or should be) used by a much broader
fraction of the population. Teaching to audiences which
comprised a fairly broad spectrum of the general public
made us realize that for beginners
coping with error messages was {\it the}
major hurdle. Many of our students, especially those who did
not have the opportunity to practice programming for days and weeks,
were unable to overcome this obstacle and got discouraged. 
In short, in the present system occasional
programmers are just left out in the cold.
\qpar
As a confirmation of the key-role of the error messages,
it can be observed that the few
programming languages in which error messages have been made 
more user-friendly quickly become very popular among students.
A case in point is the Python language.
\qpar

Clearly, the fact that the error messages are in English is only
a part of the problem. However, it adds an additional difficulty.
To get a proper understanding of this point one should keep
in mind two key-observations regarding English as a second language.
They are explained in the following subsections.

\qA{English as a second language}

Let us consider a French or Japanese tourist who visits
London or Manhattan. 
In middle- and high-school he has had English lessons 
but in the meanwhile he forgot most of what he (or she) had learned.
Thanks to his tourist guidebook he will be able to ask the
appropriate question to a police officer: \qL
``Excuse me Sir, I'm looking for the nearest post-office.''
In response, he may get the following answer.
``Well, you just have to walk across the park, the post-office
is located just after the second block on your right. It may take you
fifteen minutes at normal speed.''\qL
Probably our poor tourist will understand barely one-half
of the answer and most likely he will not be able to reach the
post-office without asking the same question again to
other persons.
\qpar

What connection does this story have with programming?
Very simple. Reading the question in the tourist guide
is similar to writing programming code in English. It does 
not require a good
knowledge of English because one needs to know
only a limited number of pre-defined words.\qL
On the contrary, the answer of the police officer is similar
to the ``answers'' generated by the compiler in the sense 
that compilers use a broad set of words that are not pre-defined
in any obvious way. In addition these words are included in
sentences. Here is an example from a Fortran compiler:\qL
``Invalid reference to variable in NAMELIST input'',
\qpar
 
Actually,
this example might well give a good reason for {\it not}
doing any translation. Indeed, almost all the words used in this
error message are exactly the same in English and in French. 
The only true English word may be ``input'' (entr\'ee) but it is
so commonly used in the pidgin form of French that is spoken 
nowadays that one can assume that it would be understood
by almost everybody. As a confirmation, let us have a look at how
Google translates the previous
message.The translation reads as follows.\qL
``R\'ef\'erence non valide \`a la variable en entr\'ee NAMELIST''.\qL
One can see that the
translation is almost identical to the English version except
for word order.
\qpar

This leads us to our second point about English as a 
second language.

\qA{Scientific vocabulary in various languages}

Between 2008 and 2014
one of us (in our research group)
has had the opportunity to teach in Beijing
and over those years
he was fairly slow to realize that whereas almost all
scientific words are the same in English and in French, in Chinese
they are completely different. An obvious implication is that
Chinese students, even those who are fairly fluent in English,
will {\it not} understand the scientific vocabulary
used in a course in physics or computer science unless they
have devoted some special time and efforts to learn it. 
As shown in the table below,
this observation about Chinese also applies to Hindi, Japanese or Korean.

\begin{table}[htb]

\small

\centerline{\bf Table 1\quad Examples of scientific words in various
languages.}

\vskip 5mm
\hrule
\vskip 0.7mm
\hrule
\vskip 2mm

$$ \matrix{
\qtb
\hbox{English} \hfill &\hbox{French} \hfill &\hbox{Spanish} \hfill &
\enskip &\hbox{Chinese} \hfill & \hbox{Hindi} \hfill &
\hbox{Japanese} \hfill & \hbox{Korean} \hfill&\hbox{Russian} \hfill \cr
\noalign{\hrule}
\qtH
\hbox{hydrogen} \hfill &\hbox{hydrog\`ene}\hfill&\hbox{hidr\'ogeno}\hfill &
\quad &\hbox{q\={\i}ng} \hfill & \hbox{h\=aidr\=ojana} \hfill &
\hbox{suiso} \hfill & \hbox{suso} \hfill&\hbox{vodorod} \hfill\cr
\hbox{invalid} \hfill &\hbox{invalide}\hfill&\hbox{inv\'alido}\hfill &
\quad &\hbox{w\'uxi\`ao} \hfill & \hbox{am\=anya} \hfill &
\hbox{muk\=o} \hfill & \hbox{yuhyo} \hfill&\hbox{nevernyi} \hfill\cr
\hbox{program} \hfill &\hbox{programme}\hfill&\hbox{programa}\hfill &
\quad &\hbox{jihu\`a} \hfill & \hbox{k\=aryakrama} \hfill &
\hbox{puroguramu} \hfill & \hbox{peulogeulaem}\hfill&\hbox{programma}\hfill\cr
\qtb
\hbox{reference} \hfill &\hbox{r\'ef\'erence}\hfill&\hbox{referencia}\hfill &
\quad &\hbox{c\=ank\u ao} \hfill & \hbox{sandarbha} \hfill &
\hbox{rifarensu} \hfill & \hbox{chamgo} \hfill&\hbox{ssylka} \hfill\cr
\noalign{\hrule}
} $$
\vskip 1.5mm
\small
Notes: In this short selection of words, Japanese is seen 
to be closer to
English than is Chinese, Hindi or Korean. Incidentally,
for a number of languages (e.g. Arabic, Hebrew, Vietnamese)
Google-translation does not give phonetic transcriptions.
That is why such languages were not included in the table.
\qL
{\it Source: Google-translate (the word for ``invalid'' in Russian
has been replaced by ``nevernyi'').}
\vskip 5mm
\hrule
\vskip 0.7mm
\hrule
\end{table}


\qI{Which method?}

How can one translate an error message
generated by a compiler? We copy the message, paste it into
the entry window of an online translator, copy the translation and
finally paste it below the initial message.
Thus, a first method would be to make this procedure automatic.
\begin{figure}[htb]
\centerline{\psfig{width=15cm,figure=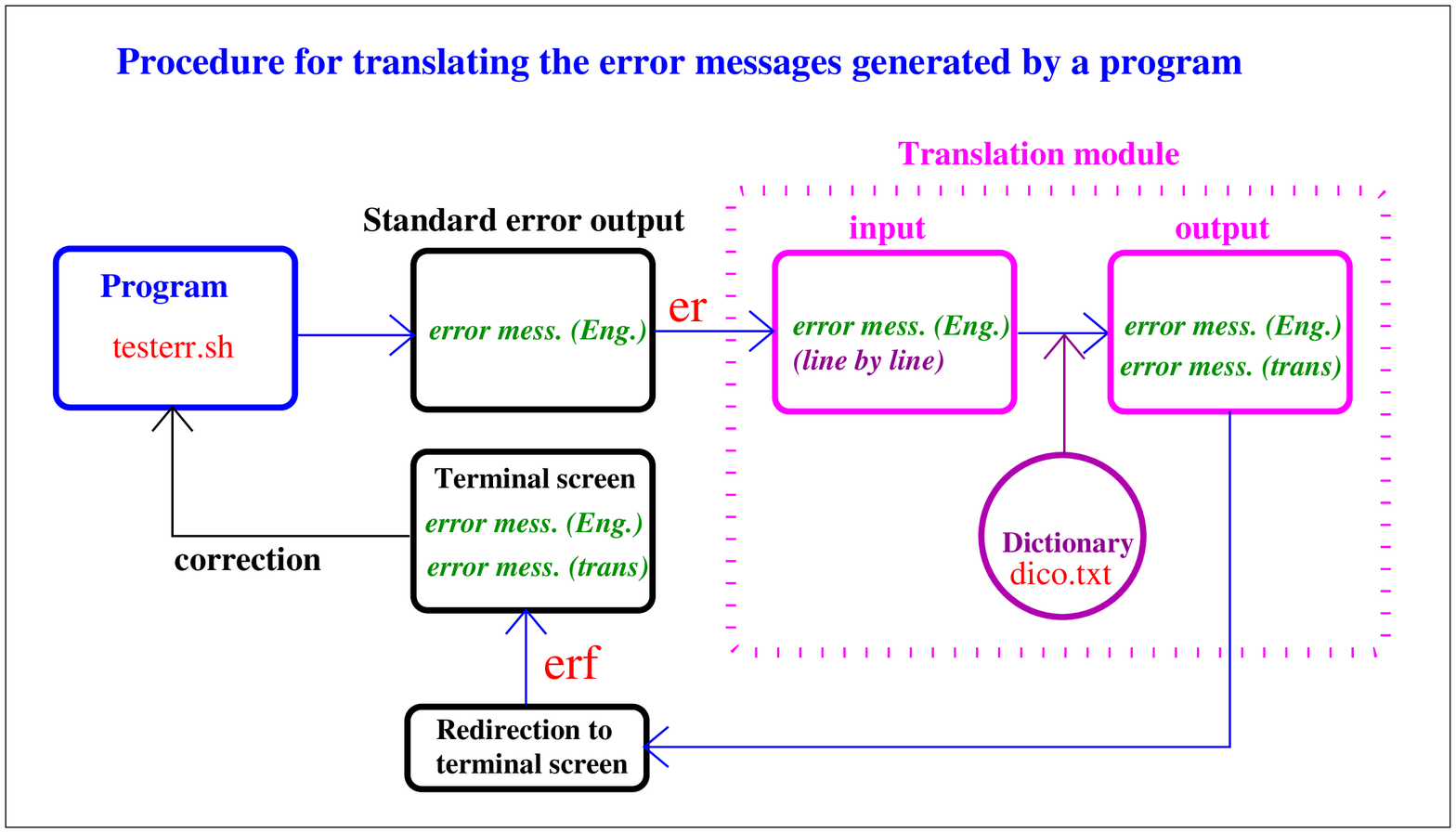}}
\qleg{Fig. \qhu 1\qhv Procedure for translating the error messages
generated in the execution of a program.}
{The words in red refer to the names of the files used in
the script dicf.sh.}
{}
\end{figure}
%
However, this method relies on an online translator
such as for instance Google-translate. Below, it will
be explained why this method is not satisfactory.
\qpar

We were told that an alternative method would be
to use ``gettext'' together with ``poedit''.
However, as we are
not familiar with this software we did not try this
option.
\qpar

Here we will present another solution
that we call the dictionary procedure in which the translation
module is simply an appropriate dictionary, 
whether self-made or web-based.

\qA{How Google-translate handles code words}

The translations generated by Google-translation have both
advantages and shortcomings.
\qpar
Among the advantages is of course the fact that they are
available for a broad range of languages.
The second advantage is that, at least for 
Japanese and European languages
(see Table 2 below)
Google-translate more or less recognizes
the words which belong to the code and does not attempt
to translate them. ``More or less'' means that 
some code words  nevertheless are translated.
As shown in Table 2 for
translations into French and German on average
about 84\% of the code
words are not translated. As an illustration, one can 
mention the case of an  ``IF $ \ldots $ 
THEN'' instruction with respect to French and German. 
In the case
of French, IF was not translated but THEN was translated
into ``Alors'', whereas in German (for exactly the
same error message) neither IF nor THEN were translated.
\qpar

For Chinese, Hindi, Korean and Russian the percentage of the code
words which were translated is much higher.
%
\begin{table}[htb]
\small
\centerline{\bf Table 2\quad Percentage of code words 
that were translated by Google-translate}

\vskip 5mm
\hrule
\vskip 0.7mm
\hrule
\vskip 2mm

$$ \matrix{
\qtb
\hbox{} \hfill &\hbox{German} &\hbox{Japanese} &
\hbox{French} & \hbox{Chinese} &
\hbox{Korean}  & \hbox{Russian} & \hbox{Hindi}\cr
\noalign{\hrule}
\qth \qtb
\hbox{Code words translated} \hfill & 9\% & 18\%&
23\% & 41\% & 50\%  & 55\% & 68\%\cr
\noalign{\hrule}
} $$
\vskip 1.5mm
\small
Notes: Ideally, none (i.e. 0\%) of the words 
belonging to the computer code should be translated. 
Actually, 
the percentages range from 9\% for German to 68\% for Hindi.
The fact that for 4 of the 7 major languages under consideration
the percentage is higher than 40\% suggests that in its present
state Google-translate
can hardly be used for translating error messages.
The data are based on a sample of words
such as NAMELIST, REWIND, OPEN, ENDFILE, VECTOR/INPUT and
so on; in order to ensure context,
they were submitted embedded within the messages
in which they occurred.
{\it }
\vskip 5mm
\hrule
\vskip 0.7mm
\hrule
\end{table}

\qA{Instability of Google's translations }

The fact (mentioned above)
that translations into French and German behave
differently is not very surprising. More puzzling is the fact that
within one language
the same sentence is translated differently when it occurs
in separate error messages. Here is an illustration for translations
into French.
In one message ``THE  OFFENDING STATEMENT IS'' is translated
into the fairly funny sentence ``LE COMPTE DE D\'ELINQUANCE EST''
whereas in another message the same words are correctly
translated into ``LA D\'ECLARATION INCRIMIN\'EE EST''. 
\qpar

This instability creates a real problem because it makes almost
impossible any attempt of introducing rules that would improve
the translations provided by Google-translate.

\qA{Replacing the translation module by a dictionary}

In Table 1 we observed that for many non-European languages
it is the scientific vocabulary which is the major difficulty.
Moreover, Table 2 showed that for these languages the translations
generated by Google-translate were not very appropriate.
Therefore, it may be tempting to replace the translation module
by a self-made lexicon. 
Because it eliminates all syntactic difficulties 
such a dictionary is easy to create. For some programming
languages (e.g. Fortran) lists of error messages are available
on the Internet. Alternatively, at least for open source languages,
the error messages may be extracted from the source code.
In addition to translations, where required, the lexicon 
may also provide short explanations.
\qpar
Although a lexicon may appear as 
a fairly rudimentary form of translation device,
it may be quite useful nevertheless. Its initial implementation
does not require much time and effort and, once implemented,
it can be improved little by little by taking into account
the observations and feedback of the students.

\appendix

\qI{Appendix A: Code for automatic word by word translation}

The code given below is independent of the language $ L $ into
which one wishes to translate the error messages.
For the purpose of illustration we will consider the case of
$ L $=French. The translations are contained in the
dictionary represented by the file \verb?dico.txt? (see below)
If one wishes to translate into another language, for
instance Japanese, one must replace this file by an English to
Japanese dictionary.
\qpar

Regarding the programming language, we consider 4 cases:
Bash, Python, Fortran and C. Because bash is just the language
of Linux (or at least one of the versions) its execution will
be the same on all Linux systems. For Python, Fortran and C there
may be some slight differences from one Linux system to another. 
\qpar

Below we give the content of the dictionary, then the code of the
scripts, then some explanations, and finally the results given
by simple tests.
\qpar

{\color{blue} Dictionary: \verb?dico.txt?}\qL
\verb?ENGLISH -> FRENCH DICTIONARY FOR ERROR MESSAGES ?\qL
\verb?USED FOR MESSAGES GENERATED BY bash,python,C,Fortran?\qL
\verb?at=a l'endroit indique ?\qL
\verb?before=avant ?\qL
\verb?call=appel (par ex. appel a une instruction) ?\qL
\verb?empty=vide ?\qL
\verb?error=erreur ?\qL
\verb?expected=attendu ?\qL
\verb?file=fichier ?\qL
\verb?forbidden=interdit ?\qL
\verb?found=trouve ?\qL
\verb?function=fonction ?\qL
\verb?in=dans ?\qL
\verb?last=en dernier ?\qL
\verb?last)=en dernier ?\qL
\verb?last):=en dernier lieu, a la fin ?\qL
\verb?length=longueur ?\qL
\verb?line=ligne ?\qL
\verb?many=nombreux ?\qL
\verb?matching=correspondant, ayant la meme forme ?\qL
\verb?most=le plus  ?\qL
\verb?(most=le plus  ?\qL
\verb?near=pres de ?\qL
\verb?number=nombre ?\qL
\verb?operand=operande (3+5: 3 et 5 sont deux operandes) ?\qL
\verb?statement=instruction, commande ?\qL
\verb?token=occurrence d'un symbole ?\qL
\verb?too=trop ?\qL
\verb?traceback=remonter a la source ?\qL
\verb?TypeError=type de l'erreur ?\qL
\verb?unexpected=non attendu ?\qL
\verb?unknown=inconnu ?\qL
\verb?unsupported=non reconnu, incompatible avec le systeme ?\qL
\verb?warning=avis (n'empeche pas l'execution)  ?\qL
\verb?without=sans  ?\qL

\qpar
Two observations are in order.
\qbu The translation can be supplemented with an explanation.
This was done for the words: call, matching, operand, 
unsupported.
\qbu The dictionary can be built little be little as the students
are confronted with new error messages. Even if it is built by the 
professor before the start of the course, one should take advantage
of the classes to make sure that the translations and explanations are
well understood. If they are not, the dictionary may be improved.

\qA{Scripts used in the execution of a bash program}

We need three kinds of scripts.
\qee{1} A test-program called \verb?testerr.sh?
which contains an error and will therefore generate
an error-message.
\qee{2} The main script called \verb?dicf.sh?
which reads the error messages and
produces the translations.
\qee{3} The main script creates a new file called \verb?erf?. 
It one runs the script a second time there will be a message
saying ``File erf already exists''. In order to avoid this
message, the file ``erf'' must be deleted before running
dicf. This is done by the script: \verb?frem.sh?
\qpar

{\color{blue} \verb?testerr.sh?}\qL
\verb?#!/bin/bash ?\qL
\verb?#BASH SCRIPT: PRODUCES OUTPUT (bonjour) AND ERROR (; ;)?\qL
\verb?# ?\qL
\verb? echo Bonjour ?\qL
\verb? echo a ; ;  echo b ?
\qpar

{\color{blue} \verb?dicf.sh?}\qL
\verb?#!/bin/bash ?\qL
\verb?# USAGE: bash dicf.sh .\testerr.sh ?\qL
\verb?#  echo ?\qL
\verb?#  ?\qL
\verb?# (1) EXECUTION + REDIRECTION OF ERROR MESSAGES TO err ?\qL 
\verb?#     ONE KEEPS ONLY THE FIRST n1 LINES OF err ?\qL
\verb?# ?\qL
\verb? echo '           RESULTATS' ?\qL
\verb? nl=10 ?\qL
{\color{red}\verb? $1 2> err ;  head -$nl err > er?}\qL
\verb?# ?\qL
\verb?#     (2) ONE READS THE ERROR MESSAGES LINE BY LINE ?\qL 
\verb?# ?\qL
\verb? echo '           ERREURS' ?\qL
\verb? li=0 ?\qL
\verb? fd=dico.txt ?\qL
\verb? while read a1 a2 a3 a4 a5 a6 a7 a8 a9 a10 a11 a12 ; do ?\qL
\verb? li=$(( $li+1 )) ; echo ; echo LIGNE $li ?\qL
\verb? echo $a1 $a2 $a3 $a5 $a6 $a7 $a8 $a9 $a10 $a11 $a12 ?\qL
\verb?# ?\qL
\verb?#     (3) ONE TRANSLATES EACH LINE WORD BY WORD ?\qL
\verb?# ?\qL
\verb? echo '           TRADUCTION' $li ?\qL
\verb? for i in $a1 $a2 $a3 $a5 $a6 $a7 $a8 $a9 $a10 $a11 $a12?
\verb? do?\qL
{\color{red}\verb? grep -iw $i $fd ?}\qL
\verb? done ?\qL
\verb? done < er ?\qL
\verb? echo ?\qL

{\color{blue} Explanations for \verb?dicf.sh?} 
\qpar

For the readers who are not familiar with the bash programming
language the following explanations may be useful.
\qbu A bash script f.sh can be executed either by \verb? bash f.sh?
or by \verb? ./f.sh?. In the second method the instruction
has  only one string. This string is given to \verb? dicf.sh ?
as an argument. 
\qbu The key-instruction is: \verb? $1 2> err?.
\verb?$1 ? reproduces the argument and therefore runs 
\verb?testerr.sh?. The number ``2'' refers to the error output.
Through \verb? > err ? this error output is redirected to the file 
\verb? err ?. 
\qbu Each of the variables  \verb? a1,a2,a3,... ? reads one of the
words in one line of the error message. 
It is the \verb?while-do-done? loop which treats all the lines.
\qbu For \verb? i=$a1? the instruction \verb?grep -iw $a1 $fd ?
searches the dictionary \verb? $fd ? for the word \verb? $a1? and
prints the line which contains the word along with its translation.
The option \verb? i ? ignores case distinction, the option
\verb? w ? selects only matches that form whole words.
\qpar

{\color{blue} \verb?frem.sh?}\qL
\verb?#!/bin/bash ?\qL
\verb?# USAGE: ./frem f ?\qL
\verb?# IF THE FILE f EXISTS IT WILL BE REMOVED ?\qL
\verb? f=$1 ?\qL
\verb?# pwd = print working directory ?\qL
\verb? p=$(pwd) ; fn=$p/$f  ?\qL
\verb?# THE ``IF'' IS A file test operator?\qL
\verb? if [ -f $fn ] ; then rm -f $fn ; fi ?\qL

\qA{How to run the scripts}

In order to execute the test file \verb?testerr.sh? and 
at the same time to translate
the error messages one should type the following instruction on
the command line:\qL
{\color{blue} 
\verb?bash frem.sh erf ; bash dicf.sh ./testerr.sh > erf ?}

\qA{Output}

One gets the following output contained in file \verb? erf?.
\qpar
{\color{blue} \verb?erf?}
\qpar
\verb?           RESULTATS  ?\qL
\verb?Bonjour ?\qL
\verb?           ERREURS ?\qL
\verb? ?\qL
\verb?LIGNE 1 ?\qL
\verb?./testerr.sh: line 6: error near unexpected token `;' ?\qL
\verb?           TRADUCTION 1 ?\qL
\verb?line=ligne ?\qL
\verb?error=erreur ?\qL
\verb?near=pres de ?\qL
\verb?unexpected=non attendu ?\qL
\verb?token=occurrence d'un symbole ?\qL
\verb? ?\qL
\verb?LIGNE 2 ?\qL
\verb?./testerr.sh: line 6: echo a ; ; echo b' ?\qL
\verb?           TRADUCTION 2 ?\qL
\verb?line=ligne ?\qL

\qA{Scripts used in the execution of programs in Python, Fortran and C}

Nothing has to be changed in \verb? dicf.sh?, only its argument
needs to be replaced by the appropriate execution orders (see below).

\qA{Python}

A possible test file containing an error is the following.
\qpar
{\color{blue}\verb?testerr.py?}\qL
\verb?print("bonjour")?\qL
\verb?1 + "2"?
\qpar

This program will be executed by the instruction:
\verb?python testerr.py?.
\qpar

It will generate the following error messages:\qL
\verb?Traceback (most recent call last):?\qL
\verb?  File "testerr.py", line 2, in <module>?\qL
\verb?    1 + "2" ?\qL
\verb?TypeError:?\qL
\verb?unsupported operand type(s) for +: 'int' and 'str'?\qL
\qpar

In order to execute the python program and at the same time
translate its error messages one should
type the following instruction in the command line%
\qfoot{For some reason the quotes around python testerr.py 
seem to look like counter-quotes, 
but they are in fact ordinary quotes '...'.}%
:\qL
{\color{blue}
\verb?bash frem.sh erf ; bash dicf.sh 'python testerr.py'>erf?}

\qA{Fortran}

A possible test-file containing an error is the following%
\qfoot{For the cases of Fortran and C programs
we consider the occurrence of
an error in the compilation. This is
for the sake of simplicity.
In case the compilation is successful
one needs to run the same procedure a second time
with the argument of ``dicf.sh'' being ``./errF''.}%
\qpar

{\color{blue}\verb?err.f90?}\qL
\verb?program tabulation?\qL
\verb?print *, "bonjour"?\qL
\verb?write *, ""?\qL
\verb?end program?\qL

This program will be executed by the instruction: \qL
\verb?gfortran err.f90 -o errF?
\qpar

It will generate the following compilation error message:\qL
\verb?err.f90:4.5: ?\qL
\verb?    write *, "" ?\qL
\verb?         1 ?\qL
\verb?    Error: Syntax error in WRITE statement at (1) ?
\qpar

In order to execute the Fortran program and at the same time
translate its error messages one should
type the following instruction in the command line:\qL
{\color{blue}
\verb?bash frem.sh erf ;?\qL
\verb?bash dicf.sh 'gfortran err.f90 -o errF'> erf?}

\qA{C}

A possible test-file containing an error is the following.
\qpar

{\color{blue} \verb?err.c?}\qL
\verb?#include <stdlib.h> ?\qL
\verb?#include <stdio.h>?\qL
\verb? ?\qL
\verb?void main(){ ?\qL
\verb?    printf("bonjour\n"); ?\qL
\verb?    printf("\n") ?\qL
\verb?} ?

\qpar
This program will be executed by the instruction:
\verb?gcc err.c -o errC? 
\qpar

It will generate the following error message:\qL
\verb?err.c: In function ‘main’:?\qL
\verb?    err.c:7:1: error: expected ‘;’ before ‘}’ token ?\qL
\qpar
In order to execute the C program and at the same time
translate its error message one should
type the following instruction in the command line:\qL
{\color{blue}
\verb?bash frem.sh erf ; bash dicf.sh 'gcc err.c -o errC'> erf?}

\qI{Appendix B: How Europe fell behind}

It is hardly an exaggeration to say that
all major programming languages in use nowadays have been
created in the United States. A few decades ago some attempts 
in this direction had been 
made in Europe (e.g. Algol or Pascal)
but they were quickly forgotten.
It is interesting to understand why in this field
European countries became less and less effective.
To get a realistic assessment one must consider
the different sectors of the software industry
one by one: software for scientific computing,
for the Internet, for graphics, images, films, translations, 
education, genealogy, and so forth and so on.
In all these sectors the domination of US companies
is overwhelming.
\qpar

In this appendix we analyze the sociology 
of software development
in order to find a clue for why Europe lags behind.
It will be seen that a large part of the problem comes
from the fact that European countries have a narrow, purely
professional, conception of ``code writing'' and
software development. That
conception is strongly connected with the fact that 
most European people, and particularly professors and students,
see 
programming as a cryptic, purely technical activity.
A more appropriate conception would be to see
``code writing'' as a
language which allows people to create a whole new world,
just as musical notation is the key for creation in 
the world of music.

\qA{The long process of software maturation}
 
For most of the languages that we are talking about (Fortran, C, Java,
Python), their development
was a long process which covered several decades.
Often it happened that
the development of a software remained so to say in hibernation for years.
That was the case for the
``ImageMagick'' software between 1990 and 1995 until
new contributors took an interest in it.
During such ``sleeping beauty'' episodes, instead of being dropped 
or frozen, the programs
continued to be maintained until eventually brought back
to life by a new team of interested people.
Needless, to say this kind of development supposes a broad 
base of programmers, not only top programmers who work
for major companies but also grass-root programmers.
In addition, during this long maturation process 
developers and programmers needed to be supported by
non-profit foundations or by government contracts.

\qA{The role played in the United States by non-profit foundations}

As examples one can mention
the ``Free Software Foundation'' and the
``ImageMagick Studio Limited Liability Company''. Both are
non-profit, tax-exempt organizations that were created some 30 years
ago. Whereas the first one is a major actor in the free software
sector, the second is a much smaller organization focused
on image processing. Although in some cases (e.g. Unix or Python)
the DARPA (Defense Advanced Research Projects Agency) was 
quite important, in a general way a myriad of foundations supported by
donors and sponsors were the key-players.

\qA{The importance of software configuration management}

Often contributions by various programmers 
to the projects harbored by such
foundations are managed through a SCM (Software Configuration
Management) software. The phase of integrating 
new code into an existing software is
also referred to as ``Software Verification and Validation''.
It is obviously a crucial step because it allows 
software development to really become a collective creation
even though the contributors do not have the
opportunity to  meet each other.
For instance, as of May 2015, ``GraphicsMagick'' (a ramification 
of ``ImageMagick'') was using a SCM called ``Mercurial''.
The website of  ``GraphicsMagick'' also shows that over the past
10 years about 40 persons contributed by writing
additional code;
most of them did so on an occasional part-time basis. 

\qA{Connection between grass-roots and top programmers}

Usually, good national football (soccer) teams exist
in countries where football is popular and where each city has
its own team of enthusiastic amateurs.
In other words, the number and enthusiasm of grass-roots players 
conditions the achievements at the top.
The same rule seems to apply to software development. 
In Europe, it seems that there is 
a very narrow base of programmers. 
From software
engineers to developers, to programmers there are many
possible definitions of the persons working in information
technology. As a result it is hardly possible to find reliable 
comparative data%
\qfoot{Typically, persons capable of writing
(and debugging) a program of some 100 lines (in any language)
may be defined as having a programmer capability. However, such
a definition is difficult to implement at a statistical level.}%
.
However, it is common knowledge that in Europe business leaders
as well as the European Commission
routinely complain about a shortage of programmers.
\qpar

Based on our teaching experience, it can be said that no
more than 10\% of the master students in physics are ``fluent'' in
programming. True, most of them have a notion of several languages but
only few practiced long enough to become really fluent.
In fact, many students are discouraged at an early stage.
The son of one of us was a case in point. During his second year
as a master student in economics he got a course in programming.
That was of course much too late in his education. In addition,
confronted to a deluge of error messages and 
a lack of support from his teachers
in this initial stage, he lost courage and decided
that definitely programming was not his cup of tea.

\qA{Consequence of a lack of interaction between top and base: the case
  of CERN}

Experiments in particle physics pose major challenges for computer
scientists because one needs to record and sort out the information
pertaining to billions of collisions between particles. 
This makes the CERN%
\qfoot{CERN is the acronym for ``Conseil Europ\'een pour la
Recherche Nucl\'eaire'' (European Council for Nuclear Research).
The organization was founded in 1954 by 12 European countries.}
a place where new software procedures are designed, implemented
and used.
Since the start of the LEP 
\qfoot{Large Electron Positron collider, to date still the most
powerful lepton collider ever built.}
project in 1983, computer scientists at CERN had to find new
ways for making the experimental data available 
to teams of collaborators worldwide.
This lead first to the development of the ``N-upple'' tool (1988)
for sharing the data describing collision events and then, in  1989,
to the introduction of the ``World Wide Web'' by Tim Berners-Lee et Robert
Cailliau.  The LHC (Large Hadron Collider) raised challenges
of even greater magnitude because of a much larger number of
collisions.
\qpar

Yet, despite the breakthroughs made at CERN very little trickled
down to the broader software community.
Nowadays who knows that the World Wide Web started at CERN?
As a matter of fact, 99\% of the tools that allowed 
step-wise improvement
of the Internet originated in the United States. This included
for instance the development of 
Unix or of the software for creating,
converting and transforming sounds, images and films.
In other words, a trickling down process would have required
a large base of persons in various fields capable of writing
code and willing to cooperate with one another.
\qpar

Even in the more narrow field of scientific research
the diffusion of the tools created at CERN was neglected.
For instance, the programming language called PAW (Physics
Analysis at Workstations) which contained the innovative
N-upple tool was frozen in 2004 and then integrated
into a specialized computation software (called ``root'') 
almost only used by particle physicists. In other words,
although this language was both powerful and pleasant to use
no attempts were made 
to put it at the disposal of users outside of the
community of particle physicists. Such a task would have been
ideally suited for a foundation. Ensuring software maintenance
and diffusion would have been a way to fully exploit
the work and efforts of the developers of PAW 
over more than 15 years.

\qA{Conclusions}

To conclude this short inquiry into the sociology of software
development one can say three things.
\qbu Rather than a top-down
process, software development should rather be
seen as a bottom-up growth process. In other words,
the achievements of top developers depend upon the interest,
capability and activity of a broad base. 
\qbu There is good reason to think that an initial 
unfriendly contact with programming languages
discourages many students, thus preventing 
the constitution of a broad base of programmers 
especially in countries whose natural languages have minimal 
overlap with English.
\qbu It is true that a number of gifted persons 
from various countries, like
Vinod Khosla from India and founder of ``Sun Microsystems'',
Guido van Rossum from
the Netherlands and creator of Python,
Linus Torvalds from Finland and a main developer of Linux,
were able to
make major contributions, but it is also true that all
three moved to California fairly early in their carrier.

\vskip 5mm
{\bf Acknowledgments}\quad The author wishes to express his
gratitude to his colleagues Marc Bellon, Vladimir Dotsenko,
Harold Erbin, Marco Picco and Hugo Ricateau
for their help and cheerfulness. He is also grateful to 
Gilles Dowek (INRIA), Christophe Gragnic 
and Hiroshi Iyetomi (University of Niigata)
for their interest in this project and their advice.

\end{document}